\documentclass[IEEE Transactions on Instrumentation and Measurement]{IEEEtran}

\usepackage[bookmarks=false]{hyperref}
\usepackage{graphicx}
\usepackage{amssymb}
\usepackage{latexsym}
\usepackage{floatrow}
\usepackage{url}
\usepackage{xcolor}
\usepackage{hyperref}
\usepackage{framed,multirow}
\usepackage[ruled,vlined]{algorithm2e}
\usepackage{booktabs}
\usepackage{multicol}
\usepackage{lipsum}
\usepackage{mwe}
\usepackage{multirow}
\usepackage{relsize}
\usepackage{float}
\usepackage{floatrow}
\usepackage{verbatim}
\usepackage{tabularx}
\usepackage{amsmath,epsfig}
\usepackage{bbm}
\usepackage{relsize}
\usepackage{booktabs}
\usepackage{amssymb}
\usepackage{latexsym}
\usepackage{upgreek}
\usepackage{cite}
\usepackage{subcaption}
\usepackage{xcolor}

\ifCLASSINFOpdf
\else
\fi
\begin{document}
%
\title{Novel Pooling-based VGG-Lite for Pneumonia and Covid-19 Detection from Imbalanced Chest X-Ray Datasets}
%
%
%

\author{Santanu Roy,
        Ashvath Suresh, Palak Sahu, and Tulika Rudra Gupta
     
\thanks{Dr. Santanu Roy and Ashvath Suresh are with the Department of Computer Science and Engineering, at Christ (Deemed to be University), Kengery Campus, Bangalore, India. Whereas, Palak Sahu is working with the Department of Computer Science and Engineering, at NIIT University, Rajasthan, India. Dr. Tulika Rudra
Gupta is working at the Dana-Farber Cancer Institute, Harvard Medical
School, USA. \\
e-mail: santanuroy35@gmail.com; ashvath.suresh@btech.christuniversity, palak.sahu20@st.niituniversity.in,
tulika@hsph.harvard.edu}
\thanks{Manuscript is under review in IEEE Transactions on Emerging Topics in Computational Intelligence.}}
%
%

\markboth{IEEE Transactions on xxxx xxxx, ~Vol.~xx, No.~x, Month~xxxx}%
{Shell \MakeLowercase{\textit{et al.}}: Bare Demo of IEEEtran.cls for IEEE Journal}
%



\maketitle
\begin{abstract}
This paper proposes a novel pooling-based VGG-Lite model in order to mitigate class imbalance issues in Chest X-Ray (CXR) datasets. Automatic Pneumonia detection from CXR images by deep learning model has emerged as a prominent and dynamic area of research, since the inception of the new Covid-19 variant in 2020. However, the standard Convolutional Neural Network (CNN) models encounter challenges associated with class imbalance, a prevalent issue found in many medical datasets. The innovations introduced in the proposed model architecture include: (I) A very lightweight CNN model, `VGG-Lite', is proposed as a base model, inspired by VGG-16 and MobileNet-V2 architecture. (II) On top of this base model, we leverage an ``Edge Enhanced Module (EEM)" through a parallel branch, consisting of a ``negative image layer", and a novel custom pooling layer ``2Max-Min Pooling". This 2Max-Min Pooling layer is entirely novel in this investigation, providing more attention to edge components within pneumonia CXR images. Thus, it works as an efficient spatial attention module (SAM). We have implemented the proposed framework on two separate CXR datasets. The first dataset is obtained from a readily available source on the internet, and the second dataset is a more challenging CXR dataset, assembled by our research team from three different sources. Experimental results reveal that our proposed framework has outperformed pre-trained CNN models, and three recent trend existing models ``Vision Transformer", ``Pooling-based Vision Transformer (PiT)'' and ``PneuNet", by substantial margins on both datasets. The proposed framework VGG-Lite with EEM, has achieved a macro average of 95\% accuracy, 97.1\% precision, 96.1\% recall, and 96.6\% F1 score on the ``Pneumonia Imbalance CXR dataset", without employing any pre-processing technique. All the codes along with their classification reports, graphs, and confusion matrices are (publicly) available on a GitHub link: 
https://github.com/dp54rs/Pneumonia-Detection-Attention-Model
\end{abstract}


%
\IEEEpeerreviewmaketitle

\section{Introduction}
\IEEEPARstart{P}neumonia, a prevalent and
potentially life-threatening
respiratory infection, encompasses
a spectrum of lung inflammations
caused by various pathogens such as
viruses, bacteria, and fungi. It causes inflammation in the air sacs and as a consequence, there is breathing difficulty and lungs are congested by dry productive cough [1]. Moreover, this is very difficult to identify whether this infection happens due to bacteria or non-bacteria (i.e., Viral Pneumonia). Lung opacity, often referred to as ground-glass opacity [2], emerges as a notable abnormality in the lungs. This ground-glass opacity is a radiological term that indicates hazy gray areas in the CXR or Computational Tomography (CT) images which generally occurs due to air sacs (in lung portions) becoming partially filled with some kind of fluid or pus. Tuberculosis (or, TB) is a distinct variant of Pneumonia, caused by Koch's bacillus bacterium [3]. Pulmonary TB occurs when the bacillus bacterium attacks the lungs, moreover, it can spread to other organs of the body [4]. Pulmonary TB is curable if it is detected in early phases. The automatic detection of Pneumonia disease has become a more popular topic recently, since the inception of Covid-19. This Covid-19 disease is another form of Pneumonia, which is caused by Severe Acute Respiratory Syndrome-2 (SARS-2) [5]. SARS is an exceptionally contagious virus and it spreads predominantly through physical contact with human beings, causing significant effects on the human immune and respiratory systems [6]. As a consequence, the mortality rate due to COVID-19 disease had increased (in 2020-21) by approximately 54\% in the USA and 10.4\% in Europe [7]. WHO already declared Covid-19 as a pandemic in March, 2020 [7]. However, in the last three years (from April 2022 to the present), the threat (of death) due to Covid-19 disease has been drastically reduced [8]. 

\begin{figure*}[h]
		\centering
		\includegraphics[width=18.2cm,height=4.1cm]{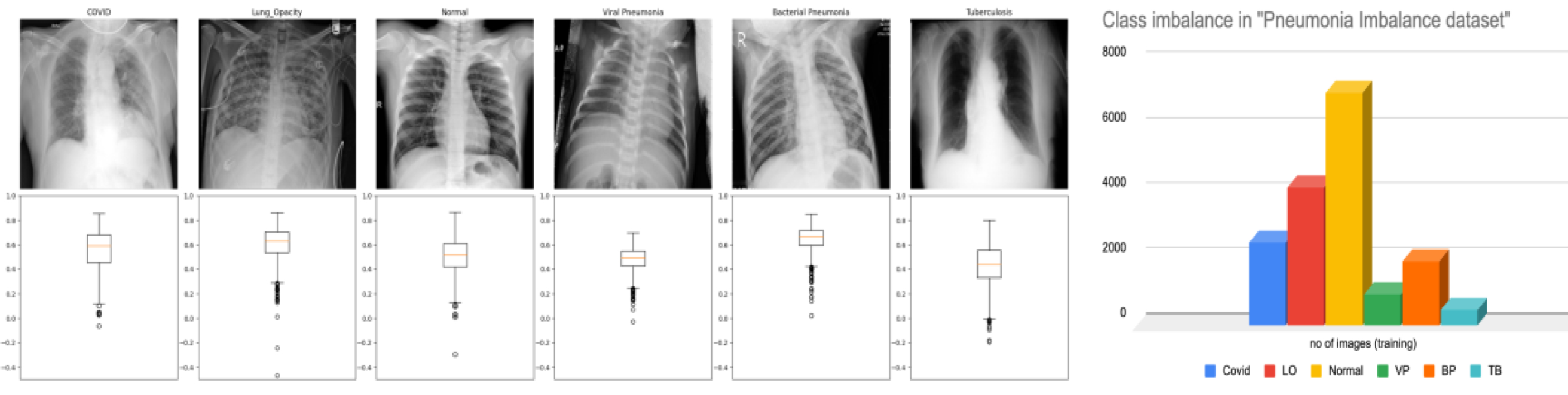}
		\caption{First image represents various classes of ``Imbalanced Pneumonia" dataset along with its box plot of correlation co-efficient, $2^{nd}$ image shows the class imbalance due to number of images differ per class}
	\end{figure*}
All the aforementioned `Pneumonia' diseases are mostly diagnosed either by Chest X-Ray (CXR) images or Computational Tomography (CT) images, since the effects of those diseases are very much prominent in the human lung system [9]. The most frequently employed Covid-19 detection technique is ``Reverse Transcriptase-Polymerase Chain Reaction (RTPCR)''[10]. However, it is a long and manual process, and is fraught with the problem of less sensitivity for Covid-19 detection.
Nevertheless, the availability of CT images' dataset is very limited and costly compared to CXR images [10]. These above-mentioned facts make the ``Chest X-Ray (CXR) images'' an automatic choice for Pneumonia and Covid-19 detection. The objective of this research is to develop an automatic computer-assisted diagnosis (CAD) system for Covid-19 and other Pneumonia variants. Additionally, distinguishing between Covid-19 patients and those with other illnesses such as Bacterial Pneumonia (BP) and Viral Pneumonia (VP), is a challenging and tedious task for medical professionals and researchers. An accurate diagnosis of the disease (whether it is VP or Covid), is crucial for the subsequent treatment of patients. Therefore, building a reliable and automated CAD system for Covid-19 and Pneumonia diseases (from Chest X-Ray images), is still a prominent research topic in the recent advent. 



The rest of the paper is organized in the following way: Section-II presents a brief description of dataset preparation. Section-III explains the existing methods of CAD and contribution of the paper. In Section-IV, the entire proposed methodology is explained with a mathematical analysis of 2Max-Min pooling technique. In section-V, experimental results and their analysis are presented. Finally, in Section-VI, we present our conclusion and future work.

\section{Dataset Preparation}
We have implemented several existing CNN models on a publicly available Chest X-Ray (CXR) dataset [11] in which there were four classes: (I) Normal, (II) Covid, (III) Lung Opacity (LO) and (IV) Viral Pneumonia (VP). This dataset is highly class-imbalanced since the number of images in different classes differs significantly. Researchers from Qatar University, and the University
of Dhaka, Bangladesh have created this dataset [11]. In this dataset, Covid class was made by three different updates which were assembled from different hospitals. Thus, Covid class has a huge intra-class (statistical) variability inside this dataset. We have also made a new CXR dataset for Pneumonia detection, by amalgamating this original CXR dataset with two other sources [12] and [13], and eventually, we have incorporated two new classes ``Tuberculosis (TB)" and ``Bacterial Pneumonia (BP)" into this new dataset. Moreover, we deliberately maintain a substantial class imbalance across various classes. In the training set, we have chosen 1946 images for BP, 2531 for Covid, 4209 for LO, 7134 for Normal, 490 for TB, and 941 for VP classes. This deliberate decision increases the challenges in the dataset compared to its previous version. We have labeled this new dataset as ``Pneumonia Imbalance CXR Dataset" and have made it accessible online via the Kaggle platform:
(https://kaggle.com/datasets/96e396523f81bfeeb0ca37ebf1501
76494cedab8ac4d97ecdec4d244175c3f24).

This is to clarify that the images from four different classes, we keep unchanged from the previous version of the CXR dataset [11]. Rather, we isolated ``Bacterial Pneumonia (BP) class" images from the first source [12] and included them as a new class directly into the new dataset. Likewise, TB images were included as a separate class from the second source [13]. Compared to the existing CXR dataset, this new dataset is more skewed and challenging, making it more representative of real-world hospital scenarios. This can be further noticed from Fig.1b the number of images per class is more diverse in this CXR dataset, thus, making the problem more challenging for conventional neural network. Later, in the results and analysis part, it is shown that those conventional models, including CNN and Vision Transformer, did not perform well on this ``Pneumonia Imbalance dataset". Moreover, Fig.1a showed the box plot diagram of correlation co-efficient [14] of various classes. Here, we take any random image from each class and compute correlation co-efficient with respect to all other images in that class and thereafter we plot these box plot diagrams. From this Fig. 1a, it is evident that the width of the box plots is greater for the Covid, Normal, and TB classes, indicating a higher intra-class variance for those classes. The significant intra-class variance [14] within these classes makes the 'Pneumonia Imbalance' dataset particularly challenging for deep learning models. Furthermore, the correlation coefficient between the two classes, presented on the Kaggle dataset link, can be interpreted as inter-class similarity [15] measure. This ``inter-class similarity" should ideally be minimal for better classification outcomes. Nonetheless, the correlation table reveals that the inclusion of Bacterial Pneumonia (BP) in the dataset increases the inter-class similarity (with a correlation of 0.67 between BP and VP, which is relatively higher than in other cases), thereby making the classification task more complex and challenging than before.

\section{Existing Methods}
Numerous researchers ([10],[16],[17], [18], [19]) have come up with novel CNN architectures to detect Covid and Pneumonia diseases efficiently from the CXR images. L. Wang et al. [10] designed a novel CNN architecture, called as Covid-Net which was designed based on a human-machine collaborative strategy. Their architecture is called Projection Expansion Projection Extension
(PEPX). One basic unit of PEPX is comprised of two 1$\times$1 convolutional layers, followed by 1 Depth-Wise Separable convolutional (DWSC) layer, followed by two 1$\times$1 convolutional layers. The 1$\times$1 filters are employed in their model to control the spectral dimension inside CNN. Despite utilizing a very lightweight CNN architecture, the PEPX model contains 183 million trainable parameters, making it prone to overfitting on limited or imbalanced CXR datasets. D. Das et al. [16] proposed a truncated Inception-Net model in which they utilized a lighter version (only four basic units) of Inception-V3 model. Moreover, they leveraged Global Average Pooling [43] instead of using Flatten layer, and avoided dense layers to reduce the trainable parameters. Kumar Aayush et al. [17] have recently introduced SARS-Net, which integrates a Convolutional Neural Network (CNN) with a Graph Convolutional Network (GCN). The authors claim that the GCN can capture relational awareness features that, when combined with the image-level features extracted by a traditional CNN, can substantially enhance the accuracy of Covid-19 detection from Chest X-Ray (CXR) datasets.

The aforementioned models tried to incorporate novelties in the CNN model architectures, in order to resolve the challenges from CXR datasets. Many researchers [20,21] attempted a direct approach, that is, either to employ Over-Sampling or Under-Sampling on the training set, or modify the Categorical Cross Entropy (CCE) loss function (into Weighted CCE [21]) in the neural network, such that it can automatically mitigate the class imbalance problem without changing the model architecture. S. Roy et al. [20] proposed a novel pre-processing method, called `SVD-CLAHE Boosting' which is comprised of both Random Under Sampling (RUS) and Oversampling. Oversampling is performed by CLAHE and one unique SVD-based image processing technique. E. Chamseddine et al. [21] proposed a framework, consisting of both SMOTE (oversampling) and weighted categorical cross entropy (WCCE) to resolve the class imbalance problem from a CXR dataset. M. Tyagi et al. [14] and S. Roy et al. [15] have tried to incorporate novelty only in terms of loss function, to eliminate the class imbalance problem from CXR dataset. M. Tyagi et al. [14] proposed a novel ``custom weighted balanced CCE (CWBCCE)" in which they assign the weights of CCE based on probabilistic notion. They derived those weights of CCE, from the statistical inference of the dataset itself. Further research on deep learning-based automatic CAD systems for Covid-19 and Pneumonia diseases can be explored in [22]-[25].

Another approach for this research is employing recent trends Vision Transformer (ViT) model [26] for the detection of pneumonia. Transformer is inspired by the concept of self supervised model [27] and it is very much popular in the field of Natural Language Processing (NLP). However, in the domain of computer vision, ViT could not yet completely replace CNN architectures [28]. This is because, unlike CNN models, ViT does not have a unique multi-scale hierarchical structure that is crucial for computer vision tasks. Moreover, heavy model of ViT makes it impractical for small medical datasets, thus, it often exhibits overfitting. Therefore, many researchers [29, 30] recently deployed novel architectures in which they integrate the basic elements from both CNN and ViT. Byeongho Heo et al. [29] proposed a Pooling-based ViT (PiT), which leverages pooling layers into the ViT model. This enhances two aspects: (I) Their model architecture will have now less the number of tokens (or, patches), thus, the number of trainable parameters has been drastically reduced. (II) It introduces a multi-scale hierarchical structure into their ViT model by utilizing pooling layers interchangeably. S. Singh et al. [31] have recently trained Vision Transformer (ViT) from scratch, for Pneumonia detection from CXR images. In contrast to CNNs, it can autonomously provide more attention to the specific fine-grained features (or local regions) that contain crucial information for Pneumonia classification. T. Wang et al. [32] introduced a hybrid deep learning framework ``PneuNet" that integrates ResNet-18 architecture with multi-head attention module (inspired by the ViT encoder). Numerous researchers [33,34] proposed ViT based hybrid models in order to address challenges in CXR datasets. Nevertheless, most of these ViTs or hybrid architectures are computationally expensive, rendering them unfeasible for very small or class-imbalanced medical datasets. Consequently, recent trend ViT or hybrid ViT approaches have not completely mitigated class imbalance issues from CXR datasets, leaving a critical research gap in the field of automated Pneumonia diagnosis. 

The other recognized attention modules invented for the image classification task include the Squeeze-and-Excitation Network (SE-Net) [35] and the Convolutional Block Attention Module (CBAM) [36]. J. Hu et al. [35] first time proposed a Squeeze Excitation Network (SE-Net). By integrating this module into a CNN model, it enables the model to better capture the inter-dependencies among channels, thus improving the model's overall performance. S. Woo et al.[36] leveraged the CBAM blocks into CNN models, combining both channel and spatial attention in order to enhance the generalization ability of the CNN model on ImageNet dataset.
B. Xiao et al. [37] proposed a Parallel Attention Module (PAM) which incorporates both channel-wise and spatial attention module (inspired by CBAM) through parallel connections with their base model in order to detect Covid-19 disease efficiently from CT images. Chiranjibi Sitaula et al. [38] have proposed an attention-based VGG-16 model in which they captured spatial relationship between pixels from CXR images, by incorporating SE-Net at deeper layers of VGG-16 model. Numerous researchers [39,40] have deployed CBAM as attention module on top of CNN models for Pneumonia detection from CXR images. 

All these aforementioned attention modules give attention either to the channels or to the spatial domain, however, these methods could not directly work in the direction of mitigating the class imbalance issue. Our proposed CEEM attention module can directly provide attention to the most important features, that are edges (of negative images). Hence, ``CEEM attention block" enables the VGG-Lite model to produce very unique kinds of ``edge enhanced and complementary features" which are essential for Pneumonia detection. This conclusion was strengthened after extensive consultations with radiologists. S.Roy et al. [41] earlier proposed a \textit{lemma} accompanied by mathematical analysis, asserting that in any CNN model, if parallel connection blocks capture distinct and salient features for classification, then it naturally alleviates the class imbalance issue to a certain degree. In another perspective, this parallel connection block works as an automatic feature augmenter, generating a greater diversity of (essential) features in the middle of the CNN model. As a consequence, the model becomes less reliant on the specific statistics of a given CXR dataset especially for minor classes, thereby enhancing the generalization ability of the model. Hence, the proposed framework mitigates the class imbalance issue to some extent. Until now, no other ``spatial attention module'' has been capable of providing direct attention to the salient features (i.e., edges), because a powerful pooling technique like ``2Max-Min pooling" had not been devised before. Hence, we believe that our proposed attention module ``CEEM" is a significant breakthrough in the field of computer vision. 

\vspace{0.1 cm}
 \textbf{Contributions of the paper:}
The contributions of this paper are presented in the following: 
\begin{enumerate}
        \item A very lightweight CNN model ``VGG-Lite", inspired by VGG-16 and MobileNet-V2, is proposed as a base model that encompasses convolutional layers, DWSC layers, and GAP layer in order to reduce the number of trainable parameters considerably.
	\item A Complementary Edge Enhanced Module (CEEM) is incorporated on top of this base model. It consists of a ``negative image" layer and a custom ``2Max-Min Pooling'' layer, followed by 3 convolutional layers. 
 \item This ``2Max-Min pooling" is an entirely novel pooling technique, introduced for the first time in this research. This pooling technique can automatically enhance some edge information inside (negative of) CXR images, thus, proposed CEEM block extracts unique kinds of features and improves the generalization ability of the model. 
  \item  In order to validate the proposed framework, it was implemented on two CXR datasets. One new ``Pneumonia Imbalance CXR  Dataset" is made and published online on the Kaggle website. 
  \item Furthermore, for checking the validity of the proposed model, a 5-fold cross-validation experiment is conducted on ``Pneumonia Imbalance dataset". 
\end{enumerate}
\section{Methodology}
The entire proposed methodology is explained in depth in the following four subsections: (a) VGG-Lite Base Model Architecture, (b) CEEM Attention block using 2Max-Min Pooling Technique, (c) Analysis of 2Max-Min Pooling with an example, (d) Some Additional Properties of the CEEM Attention Block.
\begin{figure*}[h]
		\centering
		\includegraphics[width=18.2cm,height=9.0cm]{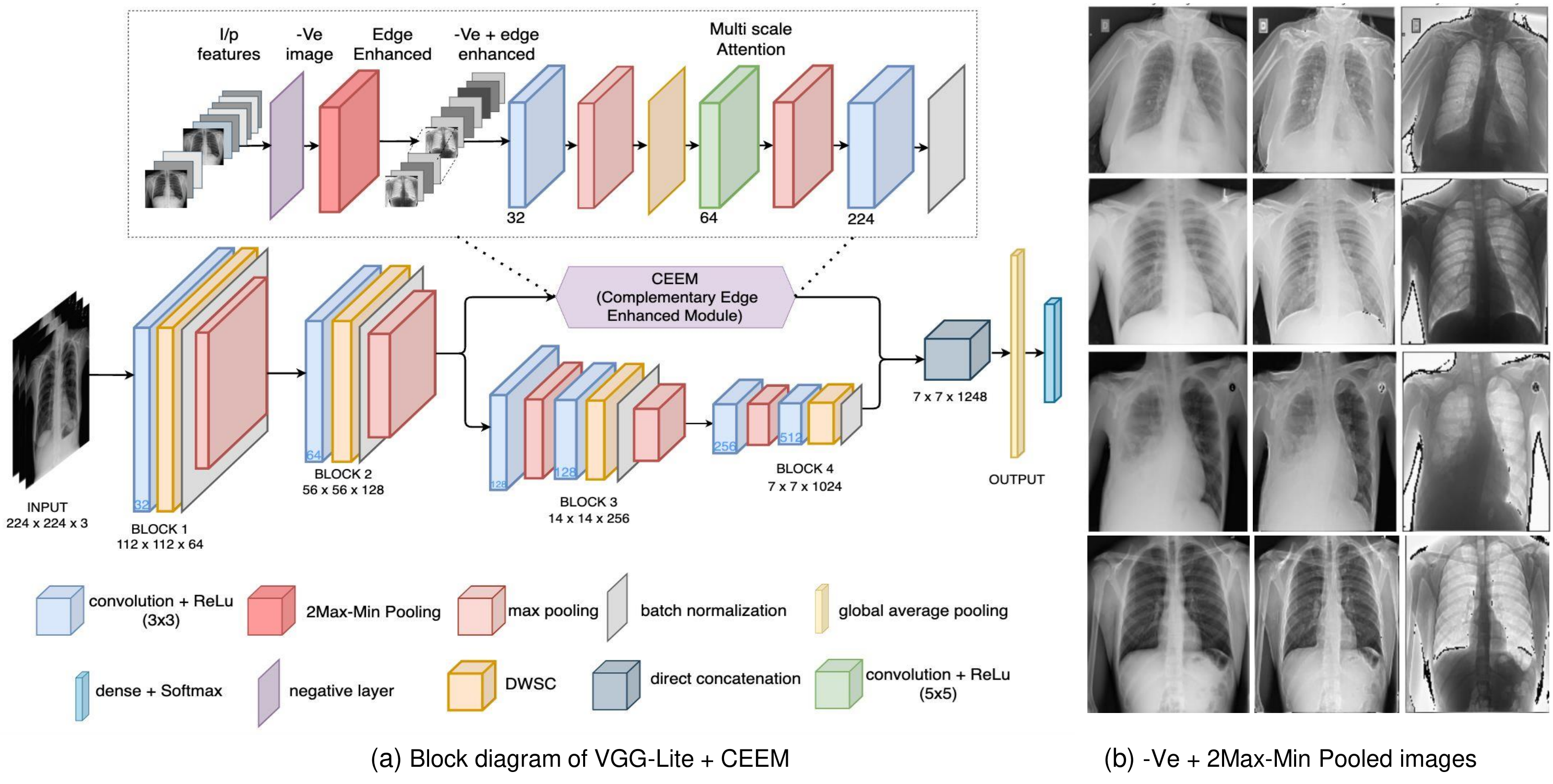}
		\caption{(a) represents block diagram of proposed framework: ``VGG-Lite"$+$ Complementary Edge Enhanced Module (CEEM). How i/p features are transformed into -Ve (complementary) and edge enhanced features by -Ve layer and 2Max-Min Pooling respectively, is demonstrated in the CEEM block. Furthermore, we present more examples of -Ve$+$2Max-Min pooled images in (b): The $1^{st}$ column represents original images, $2^{nd}$ and $3^{rd}$ column present 2max-Min pooled images and -Ve$+$2Max-Min Pooled images, respectively. \textbf{For better visualization, zooming is preferable.}}
	\end{figure*}
\subsection{VGG-Lite Base Model Architecture}
The proposed ``VGG-Lite" architecture is mainly inspired by the standard VGG-16 [42], MobileNet-V2 [43], SAM-Net [44], and a Parallel Concatenation model [41]. The proposed base model is comprised of only 6 convolutional layers, 4 Depth-Wise Separable convolutional (DWSC) layers [45], 1 GAP layer [43] and 1 output layer. In each convolutional layer and DWSC layer, the same filter kernel size (i.e., 3$\times$3) is chosen with stride=1, ReLU activation function and zero padding ``same". The structure of the proposed model is further shown in Fig.2a.

The convoluted tensor output (after $1^{st}$ convolutional layer), can be represented by the following equation.

\begin{equation}
		C_{o}(f)={ReLU({\sum_{i=1}^{p_1}{{C_i(\tau)}_{3\times3}}*{I(f)}_{m\times m}}+b)}  
\end{equation}
Here, in equation (1), `$*$' indicates convolution operation. The number of trainable parameters $h_c$ in the $1^{st}$ convolutional layer can be computed by the following equation (2).

\begin{equation}
		h_c= (3^{2}.p_0+1).p_1
\end{equation}

Here in equation (2), $p_0$ is the channel size in the previous layer (input), $p_1$ is the number of filters in the present convolutional layer, $I(f)$ is the original image having size m$\times$m, ${C_i(\tau)}$ is the convolutional filter having kernel size 3$\times$3, $b$ is the bias. 
 
On the other hand, the convoluted output tensor (after the DWSC layer [45]), in the proposed model can be represented by the following equation.

\begin{equation}
		D_{o}(f)={ReLU({\sum_{i=1}^{p_1}{{C(\tau)}.a_i}}+b)}
\end{equation}

An interesting thing to notice here in equation (3) is that ${C(\tau)}$ filter is utilized here only once per input channel, ($i=1,2,3,..$) which is followed by 1$\times$1 filter that does point-wise multiplication, thus, in equation (3), dot product of ${C(\tau)}$ and $a_i$ is considered. Hence, the number of hyper-parameters $h_D$ in the DWCS layer, in equation (4), is significantly reduced as compared to equation (2).
\begin{equation}
		h_D= (3^{2}.1+1).{p_0}=10p_0
\end{equation}
From equation (2) and (4) this is observed that, $h_c>>h_D$ for higher $p_1$. Thus, in the proposed model, in each convolutional block, at least 1 DWSC layer is employed, in order to confine the number of trainable parameters to very less value.

In each convolutional layer of the ``VGG-Lite" base model, the number of filters is chosen in the order of $2^{m}$, inspired by the original VGG-16 model [42]. In the $1^{st}$, $2^{nd}$, $3^{rd}$ and $4^{th}$ block of the ``VGG-Lite" model, the number of filters in the convolutional layer is chosen $2^5$, $2^6$, $2^7$ and $2^8,2^9$ respectively, shown in Fig.2a. Moreover, instead of using flatten layer, we have deployed Global Average Pooling (GAP) layer, inspired from MobileNet-V2 [43]. Furthermore, we avoided dense layer just before the output layer, in order to avoid over-fitting in the model. Due to the aforementioned two changes, we have been able to confine the total number of trainable parameters in the proposed ``VGG-Lite base Model" to only 2.1 million. This is considerably less than the parameters in the VGG-16 model (18 million). 

\subsection{CEEM Attention block using 2Max-Min Pooling}
 Another novelty of the proposed model is that we incorporate a Complementary Edge Enhanced Module (CEEM) into VGG-Lite base model. This CEEM block works like a spatial attention module, starting from the output of $2^{nd}$ convolutional block and directly concatenating with the GAP layer, as illustrated in Fig.2a. CEEM block consists of a custom ``negative layer", which converts the image tensor ${I_0(f)}_{w\times w}$ into its negative component, shown in equation (5). 
\begin{equation}
		I_o(f)_{w\times w}=255-{I(f)}_{w\times w}
\end{equation}
 
 Because numerous researchers [46, 47] have shown previously that negative images can extract distinct and complementary features which are essential, especially for Pneumonia detection. This is followed by a novel pooling technique, 2Max-Min pooling, with stride 2 and pooling size 3$\times$3. Any Max-pooling operation can be thought of as a function that performs down-sampling (/2) operation on the main input tensor (image) and can be represented by the following equation (6). Here, the input tensor size is reduced from $w\times w$ to $\frac{w}{2}\times \frac{w}{2}$.
\begin{equation}
		{(g_{m}{(I_o(f))}_{w\times w})}_{3\times 3|2}={I_o{(max(f)}_{3\times 3})}_{\frac{w}{2}\times \frac{w}{2}}
\end{equation}

 Now, the proposed 2Max-Min pooling can also be thought of as a function, with one Max-Pooing function and one Max-Min pooling [48] function connected in series. Therefore, 2Max-Min pooling with pool size 3$\times$3 and stride 2, can be represented by the following equation (7).
 \begin{equation*}
		{(g_{2mn}{(I_o(f))}_{w\times w})}_{3\times 3|2}={I_o{(max(f)}_{3\times3})}_{\frac{w}{2}\times \frac{w}{2}}+
\end{equation*}
\begin{equation}
		{I_o({max(f)}_{3\times3}-{min(f)}_{3\times3})}_{\frac{w}{2}\times \frac{w}{2}}
\end{equation}

Generally in digital images inside a very small (3$\times$3) patch or window, it is likely that the neighbor pixels have very similar intensity values. In other words, they do not deviate much from a particular intensity value $a_j$, even though there are edges present in the image. Thus, the edges in digital images are ramp edges. Let's assume, $a_j$ is the most frequent pixel intensity value inside a 3$\times$3 window. For simplicity of calculation, let's assume that all the pixel values inside that 3$\times$3 window can vary from $a_j-\delta_{j,-ve}$ to $a_j+\delta_{j,+ve}$, for $j^{th}$ window inside the image. This $\delta_{j,+ve}$ and $\delta_{j,-ve}$ are the maximum deviation of intensity values inside a 3$\times$3 window in the positive and negative direction respectively. Here, $a_j$ and $\delta_{j}$ are not constant values, their values may differ in each window based on the image statistics. 

Now in the conventional Max-pooling operation $g_{m}(\hspace{0.1cm})$ with 3$\times$3 window size and stride 2, let's assume that total $V$ number of windows is utilized. Maximum intensity value inside $j^{th}$ patch or 3$\times$3 window is $a_j+\delta_{j,+ve}$.
\begin{equation}
		Thus, \hspace{0.3 cm} {(g_{m}{(I_o(f))}_{w\times w})}_{3\times 3|2}={\sum_{j=1}^{V}(a_{j}+\delta_{j,+ve})}
\end{equation}
 
Similarly, in a Max-Min pooling operation $g_{mn}(\hspace{0.1cm})$, it is likely that inside $j^{th}$ patch, it will subtract the minimum value from its maximum value, thus, $a_j$ value will get approximately nullified. Additionally, 2Max-Min pooling $g_{2mn}( \hspace{0.1cm} )$, with stride 2, which is analogous to Max-Min pooling$+$Max Pooling, can be further expressed with the equation (9).

\begin{equation}
	{(g_{2mn}{(I_o(f))}_{w\times w})}_{3\times 3|2}\approx{\sum_{j=1}^{V}(a_j+\delta_{j,+ve}+\delta_{j,-ve})}
\end{equation}

Equation (9) is equivalent to one pooling operation in which total maximum deviation of intensity values ($\delta_{j,+ve}+\delta_{j,-ve}$) is superimposed with the original image inside $j^{th}$ patch. Comparing equation (8) with equation (9), it can be concluded that the operation of 2Max-Min pooling is entirely different than Max-pooling. Indeed it does a good approximation than Max-pooling, with adding (or, preserving) the extra intensity variation ($\delta_{j,-ve}$). In other words, this 2Max-Min pooling operation is analogous to producing a sharpened image or edge-enhanced features. This can be further verified in Fig.2b ($2^{nd}$ column). 

As shown in Fig. 2a, the proposed `CEEM attention block' begins with a -ve layer and 2Max-Min pooling layer to capture `Complementary' and `Edge Enhanced features' respectively. This can be further verified by the third column in Fig.2b. Hence, this attention block has been able to capture very unique kinds of features, and it is called ``Complementary and Edge Enhanced Module (CEEM)". These two layers are followed by 32 convolutional filters, a DWSC layer, 64 number of 5$\times$5 convolutional filters (to extract multiscale features [36]). Finally a convolutional layer with with 224 filters (number is chosen empirically) is applied to strengthen the impact of this spatial attention mechanism.

\subsection{Analysis of 2Max-Min pooling with an example}
 In this subsection, we have further analyzed 2Max-Min pooling by giving one simple example. We have assumed here only 1 case, we consider an edgy region (having ramp edge) of an image.

Let's assume a small portion of an image is $I_0(x,y)$ of size 7$\times$7, associated with edge portion, given in equation (10). 

\begin{equation}
I_0(x,y)=\begin{bmatrix}
109 & 111 & 111 & 110 & 111 & 112 & 112\\
113 & 114 & 116 & 115 & 117 & 115 & 117\\
119 & 118 & 117 & 120 & 120 & 122 & 151\\
135 & 128 & 127 & 126 & 128 & 130 & 161\\
143 & 142 & 142 & 141 & 139 & 142 & 142\\
158 & 157 & 154 & 157 & 151 & 154 & 151\\
170 & 169 & 165 & 165 & 162 & 163 & 160\\
\end{bmatrix}
\end{equation}

It can be observed that these intensity values in equation (10), are increasing gradually from 109 to 170, since it is likely to be a ramp edge in case of a digital image. The most frequent pixel intensity $a_j$ in the first 3$\times$3 window of the matrix is $111$ and the maximum intensity value in this 3$\times$3 window is $119$. Now if we apply 2Max-Min pooling in a local window (L, i.e., 3$\times$3) in this matrix, given in equation (10), then we shall get a 3$\times$3 matrix because stride is 2, the first element of this matrix will be $(2*maxvalue-minvalue)$.
\begin{equation}
 1^{st} \hspace{0.1 cm} elem  \hspace{0.1 cm} of \hspace{0.1 cm} {(g_{2mn}{(I_0(x,y))})}_{3\times 3|2}=(119*2-109)=129
\end{equation}

Equation (11) can also be analyzed by the mathematical derivations done in Section IV-B: With respect to the most frequent pixel intensity (of $a_j=111$), there will be intensity variation in a negative direction (i.e., $111-109=2$), as well as in a positive direction (i.e., $119-111=8$). Therefore, according to the equation (9), the total intensity variation can be represented by 2Max-Min pooling operation as follows. 

\begin{equation*}
 {(g_{2mn}{(I_0(x,y)_L)})}_{3\times 3|2}=Max({(I_0(x,y)_L)})_{3\times 3|2} +
\end{equation*}
\begin{equation}
(\delta_{j,+ve}+\delta_{j,-ve})=119+8+2=129
\end{equation}
Eventually, we shall get the output matrix of size 3$\times$3, which is given in the following equation (13).
\begin{equation}
g_{2mn}({I_0(x,y)})_{3\times 3}=\begin{bmatrix}
129 & 130 & 132\\
169 & 165 & 164\\
198 & 188 & 187\\
\end{bmatrix}
\end{equation}
Whereas, conventional Max-pooling operation with 3$\times$3 pool size along with stride 2, is presented below in equation (14).  

\begin{equation}
g_{m}({I_0(x,y)})_{3\times 3}=\begin{bmatrix}
119 & 120 & 122\\
143 & 142 & 142\\
170 & 165 & 163\\
\end{bmatrix}
\end{equation}
 
By comparing equations (13) and (14), it can be concluded that the result of the proposed 2Max-Min pooling is slightly different than Max-pooling. It can be observed that in case of 2Max-Min pooling, some extra information (or, small value) is added. This happens because the edge features are superimposed on the Max-pooled images. Thus, if there is greater intensity variation (or stronger edges present), then it will further increase the value inside the 2Max-Min pooled image matrix. This analysis supports the previously outlined theory and also substantiate the mathematical analysis presented in the preceding subsection.

\subsection{Some Additional Properties of the CEEM Attention Block}	
Some additional important properties of the proposed CEEM attention block are presented in the following:
\begin{enumerate}
        \item Compared to the conventional Max-pooling operation, the proposed 2Max-Min pooling operation is quite different, and it superimposes some edge (or, boundary) information on top of -ve of the original image. Hence, ``CEEM block" produces very unique kinds of ``complementary and edge-enhanced features". Consequently, CEEM block provides extra valuable information to the model and it enables the base model to converge faster to the least loss than other models, during training phase. Therefore, the proposed model will now have slightly extra time to pay attention to the minor classes, and it naturally alleviates the class imbalance issue to some extent.
        \item In contrast to the conventional SAM that mostly diminishes other channels to single channel (i.e., feature map $F_m\in R^{1\times X \times Y}$), our proposed attention block exploits multiple channels ($F_m\in R^{m\times X \times Y}$, where $m=224$), in order to enhance the effects of this spatial attention [44]. This approach eliminates the need of repeatedly applying the same SAM block within the CNN model. Instead, applying it just once while amplifying the attention effect through multiple filters is a highly effective method to boost the efficacy of the model. 
        \item Additionally, the proposed CEEM attention block captures multiscale features (similar to the SAM [44]) by incorporating 5$\times$5 convolutional filters. As a consequence, CEEM block enhances the base model's ability to extract a more global kinds of features.
        \item However, one limitation of this CEEM attention block is that, due to capturing extra intensity deviation inside the images (by 2Max-Min pooling), it may also retain noise since these variations could be caused by noise, thereby making it slightly susceptible to noise. 
        \item In order to overcome the aforementioned problem (in point 4), we have deployed the proposed CEEM along with 2Max-Min pooling technique, through a parallel branch only, but not in the main base model. The spatial attention weightage depends on the ratio of filters in the base model to those in the CEEM. We empirically set this ratio to $1024:224$, that is, roughly $4.5:1$. This approach ensures that the main base model predominantly influences the overall classification decision, while the CEEM just acts as a catalyst. In this way, the limitation of 2Max-Min pooling is mitigated in the proposed framework. 
\end{enumerate}

\section{Results and Analysis}
The Results and Analysis section can be summarized in three major parts: (A) Training Specifications, (B) Performance Comparisons and Analysis, (C) Validity checking by 5-fold cross-validation experiment. For the performance evaluation of these models, we employed ``precision",``recall", and ``F1 score" along with ``accuracy", in order to check whether the model performance is heavily influenced by class imbalance problem or not. 
 \subsection{Training Specifications:}
All of the CNN models have been built using Keras sequential API. Tesla P100 GPU was provided by Google Colab Pro service. 
\vspace{0.1 cm}
	The following training specifications are followed for overall all the existing CNN models. 
	
\begin{itemize}
		\item Original dataset is divided with 70\%-20\%-10\% ratio for training, testing and validation respectively. This splitting is done in a stratified way, which is feasible for class imbalance problem.
		\item Adams-optimizer is employed as the preferred choice of optimizer for all the models. 
		\item A batch size of 16 is utilized while training all the CNN models. 
        \item All the images are resized into 224 x 224 before feeding them into all the CNN models. 
		\item For training the standard pre-trained CNN models, we have chosen a fixed learning rate of $1e^{-4}$, with early stopping, monitored on the validation loss for 5 epochs patience. 
        \item The aforementioned five points are generally standard practices and widely adopted by numerous scientists worldwide for pre-trained CNN models. We have also trained the same with variety of batch sizes (32,12,8) and found 16 is the optimal batch size for this dataset. 
        \item However, for training the other existing CNN models (train-from-scratch), a total of 25 epochs with fixed learning rate $1e^{-4}$ was employed. Here early stopping has been avoided because of the higher complexity of these models `early stopping' often stops the training too early. 
        \item For the proposed model, we have empirically chosen an adaptive learning method, in which we train initially with a fixed learning rate $0.75e^{-4}$ (for $\frac{1}{3}$ of total 25 epochs, i.e., 8 epochs) and thereafter, adaptive learning rate ($lr=lr*0.96$) is incorporated for the rest of the epochs. This is to clarify that we had to change the learning rate a bit (from $1e^{-4}$ to $0.75e^{-4}$), because by several experiments we found that the above mentioned adaptive lr will be more effective only at this learning rate.
        \item In all pre-trained CNN models and trained from scratch models, no FC layer is taken into account, for fair comparison with the proposed model and for avoiding overfitting.
        \item For all the models, Categorical Cross Entropy is employed as a loss function. 
        \item No manual feature extraction is performed, in any of the existing CNN models / ViT frameworks, to obtain fair comparisons of the models.  
        \item No data augmentation or other pre-processing method is deployed in any of the CNN model/ ViT framework.	
\end{itemize}

\begin{table*}[t]
		\begin{center}		\caption{Comparison of class-wise classification reports (on testing), of numerous existing CNN models (train-from-scratch) with the proposed ``VGG-Lite", ``VGG-Lite$+$CEEM" models. Implementation is done on ``Pneumonia Imbalance dataset"}
		\resizebox{1.01\columnwidth}{!}{

\begin{tabular}{|c|ccc|ccc|ccc|ccc|}
\hline
\multirow{2}{*}{Classes}                               & \multicolumn{3}{c|}{PneuNet}                                            & \multicolumn{3}{c|}{Vision Transformer (ViT)}                          & \multicolumn{3}{c|}{VGG-Lite Model}                                     & \multicolumn{3}{c|}{\textbf{VGG-Lite + CEEM}}                                               \\ \cline{2-13} 
                                                       & \multicolumn{1}{c|}{Precision} & \multicolumn{1}{c|}{Recall} & F1-score & \multicolumn{1}{c|}{Precision} & \multicolumn{1}{c|}{Recall} & F1score & \multicolumn{1}{c|}{Precision} & \multicolumn{1}{c|}{Recall} & F1-score & \multicolumn{1}{c|}{Precision}      & \multicolumn{1}{l|}{Recall}         & F1-score       \\ \hline
BP                                                     & \multicolumn{1}{c|}{0.644}     & \multicolumn{1}{c|}{0.973}  & 0.775    & \multicolumn{1}{c|}{0.823}     & \multicolumn{1}{c|}{0.802}  & 0.812   & \multicolumn{1}{c|}{0.996}     & \multicolumn{1}{c|}{0.991}  & 0.993    & \multicolumn{1}{c|}{0.998}          & \multicolumn{1}{c|}{0.998}          & 0.998          \\ \hline
Covid                                                  & \multicolumn{1}{c|}{0.814}     & \multicolumn{1}{c|}{0.797}  & 0.806    & \multicolumn{1}{c|}{0.941}     & \multicolumn{1}{c|}{0.881}  & 0.910   & \multicolumn{1}{c|}{0.923}     & \multicolumn{1}{c|}{0.947}  & 0.935    & \multicolumn{1}{c|}{0.983}          & \multicolumn{1}{c|}{0.964}          & 0.973          \\ \hline
LO                                                     & \multicolumn{1}{c|}{0.896}     & \multicolumn{1}{c|}{0.679}  & 0.772    & \multicolumn{1}{c|}{0.803}     & \multicolumn{1}{c|}{0.838}  & 0.820   & \multicolumn{1}{c|}{0.906}     & \multicolumn{1}{c|}{0.909}  & 0.907    & \multicolumn{1}{c|}{0.946}          & \multicolumn{1}{c|}{0.871}          & 0.907          \\ \hline
Normal                                                 & \multicolumn{1}{c|}{0.794}     & \multicolumn{1}{c|}{0.915}  & 0.850    & \multicolumn{1}{c|}{0.890}     & \multicolumn{1}{c|}{0.881}  & 0.886   & \multicolumn{1}{c|}{0.948}     & \multicolumn{1}{c|}{0.900}  & 0.923    & \multicolumn{1}{c|}{0.922}          & \multicolumn{1}{c|}{0.973}          & 0.947          \\ \hline
TB                                                     & \multicolumn{1}{c|}{0.918}     & \multicolumn{1}{c|}{0.803}  & 0.857    & \multicolumn{1}{c|}{0.929}     & \multicolumn{1}{c|}{0.942}  & 0.936   & \multicolumn{1}{c|}{0.951}     & \multicolumn{1}{c|}{0.971}  & 0.961    & \multicolumn{1}{c|}{1.0}            & \multicolumn{1}{c|}{0.985}          & 0.992          \\ \hline
VP                                                     & \multicolumn{1}{c|}{0.000}     & \multicolumn{1}{c|}{0.000}  & 0.000    & \multicolumn{1}{c|}{0.570}     & \multicolumn{1}{c|}{0.629}  & 0.598   & \multicolumn{1}{c|}{0.763}     & \multicolumn{1}{c|}{0.996}  & 0.864    & \multicolumn{1}{c|}{0.981}          & \multicolumn{1}{c|}{0.977}          & 0.979          \\ \hline
Macro-Avg                                              & \multicolumn{1}{c|}{0.678}     & \multicolumn{1}{c|}{0.694}  & 0.677    & \multicolumn{1}{c|}{0.826}     & \multicolumn{1}{c|}{0.829}  & 0.827   & \multicolumn{1}{c|}{0.914}     & \multicolumn{1}{c|}{0.952}  & 0.930    & \multicolumn{1}{c|}{\textbf{0.971}} & \multicolumn{1}{c|}{\textbf{0.961}} & \textbf{0.966} \\ \hline
\begin{tabular}[c]{@{}c@{}}Weight-Avg\end{tabular} & \multicolumn{1}{c|}{0.765}     & \multicolumn{1}{c|}{0.793}  & 0.770    & \multicolumn{1}{c|}{0.853}     & \multicolumn{1}{c|}{0.850}  & 0.851   & \multicolumn{1}{c|}{0.930}     & \multicolumn{1}{c|}{0.926}  & 0.927    & \multicolumn{1}{c|}{\textbf{0.951}} & \multicolumn{1}{c|}{\textbf{0.950}} & \textbf{0.950} \\ \hline
\end{tabular}
}
\end{center}
\end{table*}

\begin{figure*}[h]
		\centering
		\includegraphics[width=17.9cm,height=4.7cm]{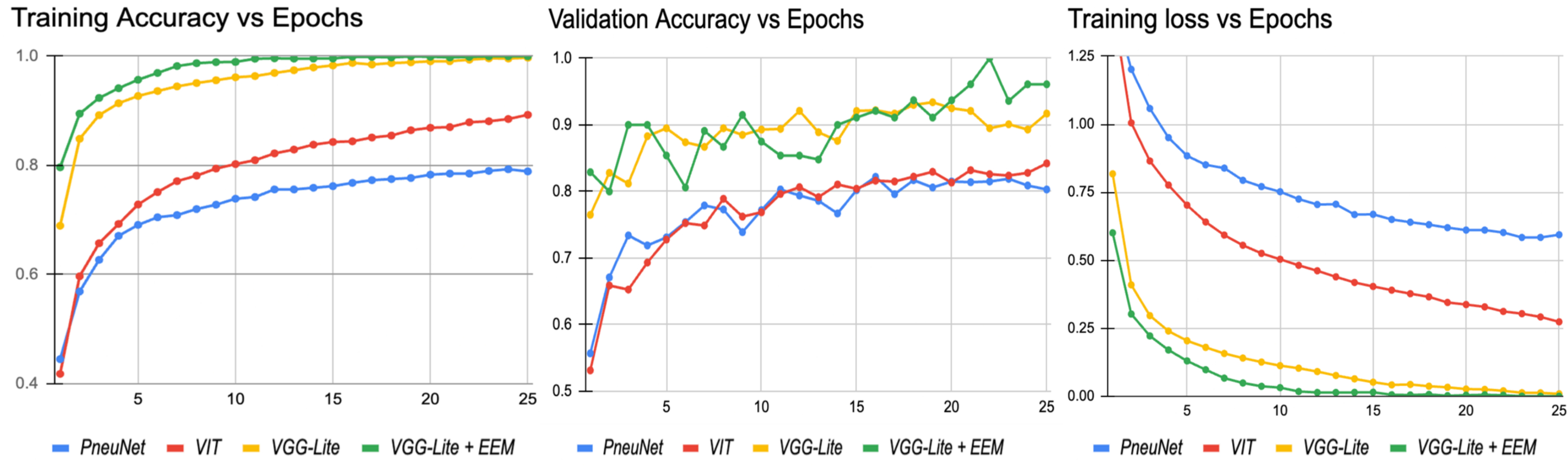}
		\caption{From left to right: Training graphs of accuracy vs epochs, validation graph of accuracy vs epochs, training graph of loss vs epochs for several models, on ``Pneumonia Imbalance Dataset''. (The validation loss vs. epochs graph is omitted here due to considerable fluctuations and limited space in the paper).}
	\end{figure*}
\begin{figure*}[h]
		\centering
		\includegraphics[width=17.9cm,height=5cm]{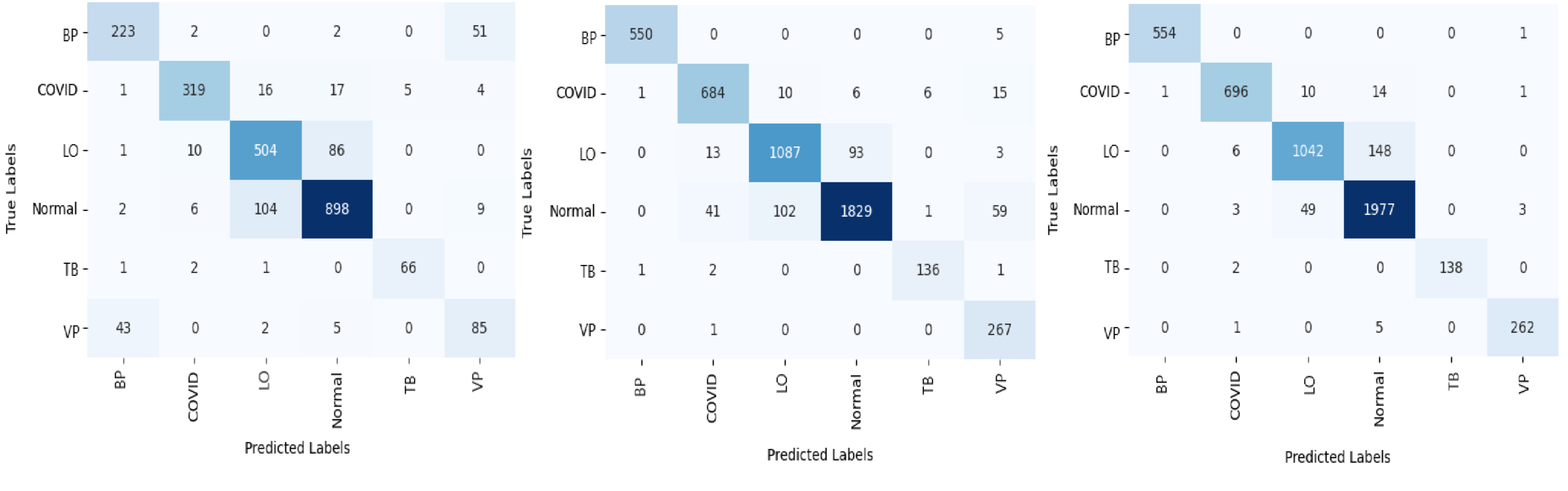}
		\caption{From left to right: Confusion matrices of Vision Transformer (ViT), Proposed VGG-Lite without attention, and Proposed VGG-Lite$+$CEEM, on ``Pneumonia Imbalance Dataset''}
	\end{figure*}

\subsection{Performance Comparison and Analysis}
This sub-section can be further divided into two parts: (I) The class-wise classification report of the proposed VGG-Lite (both with and without the attention block ``CEEM") is presented in TABLE-I. This helps us to visualize the impact of the class imbalance issue on the model. Additionally, the results of two other existing models ``PneuNet" and ``Vision Transformer (ViT)" (trained-from-scratch) are included for comparison in the same table. Moreover, the training and validation performance of these models is illustrated using comparison graphs, depicted in Fig.3. (II) Subsequently, the proposed framework ``VGG-Lite$+$CEEM" is compared with other existing models (trained from scratch), as well as other pretrained CNN models, given in TABLE-II and TABLE-III. Both these tables present the weighted average of performance metrics ``accuracy", ``precision", and ``F1 score". Additionally, one more metric ``secs/epoch" is introduced into both tables to indicate the average time (in secs) taken by the model per epoch during its training phase. Hence, this metric ``secs/epoch" is analogous to the model's time complexity. Furthermore, the graphs of all metrics vs epochs, classification reports, and confusion matrices of all these models can be found in the following GitHub link: \textbf{(https://github.com/dp54rs/Pneumonia-Detection-Attention-Model)}



 \begin{table*}[tb]
\begin{center}
\caption{Comparisons of numerous existing CNN models (\textbf{trained from scratch}) with the proposed framework (on testing). Implementation is done on both the CXR datasets (\textbf{Weighted Average} is displayed)}
\label{tab:my-table}
\resizebox{0.97\columnwidth}{!}{
\begin{tabular}{|c|cccc|ccccc|}
\hline
\multirow{2}{*}{\begin{tabular}[c]{@{}c@{}}Models\\ \end{tabular}} & \multicolumn{4}{c|}{Original CXR-Dataset} & \multicolumn{5}{c|}{Pneumonia Imbalance Dataset} \\ \cline{2-10} 
 & \multicolumn{1}{c|}{Accuracy} & \multicolumn{1}{c|}{Precision} & \multicolumn{1}{c|}{F1 score} & secs/ ep & \multicolumn{1}{l|}{Accuracy} & \multicolumn{1}{l|}{Precision} & \multicolumn{1}{l|}{F1 score} & \multicolumn{1}{l|}{secs/ ep} & \begin{tabular}[c]{@{}c@{}}Params(M) \end{tabular} \\ \hline
\begin{tabular}[c]{@{}c@{}}Covid-Net [10] \end{tabular} & \multicolumn{1}{c|}{0.898} & \multicolumn{1}{c|}{0.898} & \multicolumn{1}{c|}{0.897} & 404 & \multicolumn{1}{c|}{0.885} & \multicolumn{1}{c|}{0.885} & \multicolumn{1}{c|}{0.885} & \multicolumn{1}{c|}{480} & 183 \\ \hline
\begin{tabular}[c]{@{}c@{}} Vision Transformer (ViT) [31]\end{tabular} & \multicolumn{1}{c|}{0.843} & \multicolumn{1}{c|}{0.852} & \multicolumn{1}{c|}{0.841} & 321 & \multicolumn{1}{c|}{0.849} & \multicolumn{1}{c|}{0.853} & \multicolumn{1}{c|}{0.851} & \multicolumn{1}{c|}{382} & 85.8 \\ \hline
\begin{tabular}[c]{@{}c@{}}Pneu-Net (ResNet-18$+$ViT) [32]\end{tabular} & \multicolumn{1}{c|}{0.873} & \multicolumn{1}{c|}{0.877} & \multicolumn{1}{c|}{0.873} & 64 & \multicolumn{1}{c|}{0.793} & \multicolumn{1}{c|}{0.765} & \multicolumn{1}{c|}{0.770} & \multicolumn{1}{c|}{84} & 72 \\ \hline
\begin{tabular}[c]{@{}c@{}}Attention-based VGG-16 [38]\end{tabular} & \multicolumn{1}{c|}{0.902} & \multicolumn{1}{c|}{0.903} & \multicolumn{1}{c|}{0.902} & 104 & \multicolumn{1}{c|}{0.474} & \multicolumn{1}{c|}{0.645} & \multicolumn{1}{c|}{0.279} & \multicolumn{1}{c|}{128} & 33.3 \\ \hline
\begin{tabular}[c]{@{}c@{}}Proposed VGG-Lite \end{tabular} & \multicolumn{1}{c|}{0.927} & \multicolumn{1}{c|}{0.927} & \multicolumn{1}{c|}{0.927} & 71 & \multicolumn{1}{c|}{0.926} & \multicolumn{1}{c|}{0.930} & \multicolumn{1}{c|}{0.927} & \multicolumn{1}{c|}{99} & \textbf{2.12} \\ \hline
\textbf{\begin{tabular}[c]{@{}c@{}}Proposed VGG-Lite$+$CEEM\end{tabular}} & \multicolumn{1}{c|}{\textbf{0.943}} & \multicolumn{1}{c|}{\textbf{0.943}} & \multicolumn{1}{c|}{\textbf{0.943}} & {80} & \multicolumn{1}{c|}{\textbf{0.950}} & \multicolumn{1}{c|}{\textbf{0.951}} & \multicolumn{1}{c|}{\textbf{0.950}} & \multicolumn{1}{c|}{116} & 2.40 \\ \hline
\end{tabular}
}
\end{center}
\end{table*}

\begin{figure*}[h]
		\centering
		\includegraphics[width=17.8cm,height=3.0cm]{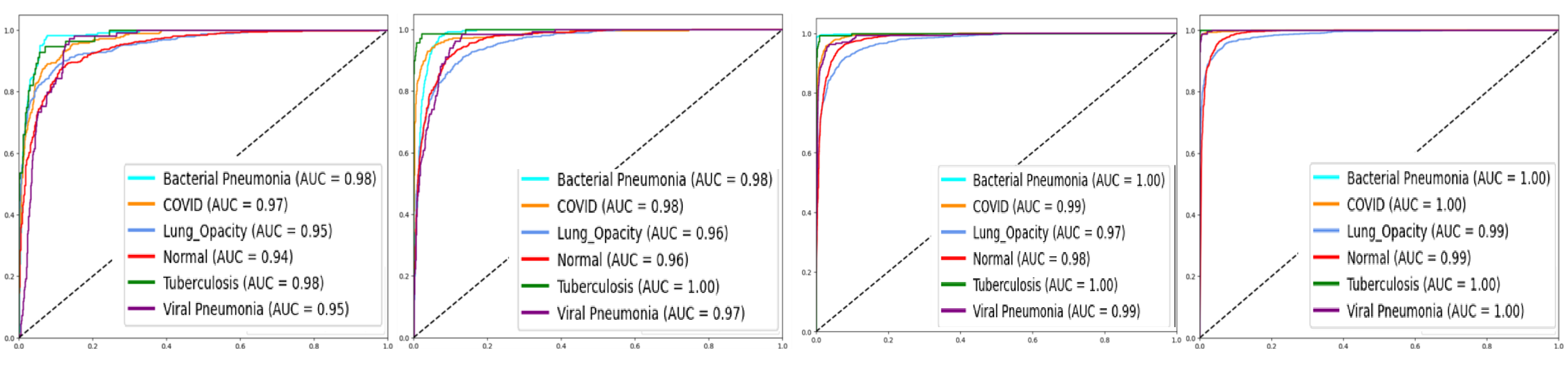}
		\caption{From left to right: RoC graph (True positive rate vs False positive rate) of the PneuNet model, ViT model, proposed model without attention, proposed model with attention respectively, on ``Pneumonia imbalance dataset". Zooming is preferable. }
	\end{figure*}
From TABLE-I, it is observed that the efficacy of the VGG-Lite Base model (without attention block), is slightly affected due to the class imbalance problem. As a result, classes with fewer training samples, such as Viral Pneumonia, exhibit lower precision of 76.3\%, shown in TABLE-I. The same can be observed for other existing trained-from-scratch models ``PneuNet" and ``ViT" as well. TABLE-I reveals that the minor class VP has not been trained well for these two existing models, resulting in very lower precision scores, with a recorded value of $0\%$, and $57\%$ for PneuNet, and ViT models respectively. On the other hand, it is evident from TABLE-I that the efficacy of the proposed framework has been considerably boosted after incorporating ``CEEM attention block''. The precision of the VP class and Tuberculosis class, have been improved to 21.8\% (from 76.3\% to 98.1\%) and 5\% (from 95\% to 100\%) respectively. These are notable improvements. From these results, it is apparent that the proposed framework mitigated the class imbalance problem substantially from the `Pneumonia Imbalance dataset'.

Moreover, from the graph in Fig.3, it is observed that the VGG-Lite$+$CEEM framework converges to the lowest loss or highest accuracy much faster than the base model (``VGG-Lite"), and other existing models during training. This validates our earlier assertion in Section IV-D. From Fig.3 it can also be observed that overall PneuNet and ViT (trained-from-scratch) models struggled to learn during the training phase itself, and both of these models' training and validation accuracy is below par compared to the proposed model. The confusion matrix (of testing set) presented in Fig.4, enables us to analyze the prediction results in a matrix form. From Fig.4, it is evident that ViT could not perform well on ``Pneumonia Imbalance dataset", as many images from BP class have been miss classified into VP class and vice-versa. This happened due to the higher inter-class similarity between class BP and VP, which was mentioned earlier in Section-II. The proposed base model ``VGG-Lite" has overcome this challenge to some extent, however, it could not distinguish ``Normal class" and ``VP class" efficiently. This was finally resolved by the proposed framework ``VGG-Lite$+$CEEM, as illustrated in Fig.4. Hence, it can be concluded that the proposed framework has not only mitigated the class imbalance issue, but also, it alleviated other challenges of datasets, such as, ``higher inter-class similarity between two classes".

 \begin{table*}[tb]
\begin{center}
\caption{Comparisons of several standard \textbf{pre-trained CNN} models with the proposed framework (on testing). Implementation is done on both the CXR datasets (\textbf{Weighted Average} is displayed)}
\label{tab:my-table}
\resizebox{0.99\columnwidth}{!}{
\begin{tabular}{|c|cccc|ccccc|}
\hline
\multirow{2}{*}{\begin{tabular}[c]{@{}c@{}}Models/\\ Methods\end{tabular}} & \multicolumn{4}{c|}{Original CXR-Dataset} & \multicolumn{5}{c|}{Pneumonia Imbalance Dataset} \\ \cline{2-10} 
 & \multicolumn{1}{c|}{Accuracy} & \multicolumn{1}{c|}{Precision} & \multicolumn{1}{c|}{F1 score} & secs/ep & \multicolumn{1}{l|}{Accuracy} & \multicolumn{1}{l|}{Precision} & \multicolumn{1}{l|}{F1 score} & \multicolumn{1}{l|}{secs/ep} & \begin{tabular}[c]{@{}c@{}}Params(M) \end{tabular} \\ \hline
\begin{tabular}[c]{@{}c@{}}VGG-16 (pre-trained)\end{tabular} & \multicolumn{1}{c|}{0.924} & \multicolumn{1}{c|}{0.925} & \multicolumn{1}{c|}{0.924} & 116 & \multicolumn{1}{c|}{0.894} & \multicolumn{1}{c|}{0.900} & \multicolumn{1}{c|}{0.893} & \multicolumn{1}{c|}{142} & 21.1 \\ \hline
\begin{tabular}[c]{@{}c@{}}Inception-V3 (pre-trained)\end{tabular} & \multicolumn{1}{c|}{0.931} & \multicolumn{1}{c|}{0.932} & \multicolumn{1}{c|}{0.931} & 122 & \multicolumn{1}{c|}{0.930} & \multicolumn{1}{c|}{0.931} & \multicolumn{1}{c|}{0.930} & \multicolumn{1}{c|}{156} & 22.3 \\ \hline
\begin{tabular}[c]{@{}c@{}}DenseNet-121 (pre-trained) \end{tabular} & \multicolumn{1}{c|}{0.885} & \multicolumn{1}{c|}{0.891} & \multicolumn{1}{c|}{0.886} & 167 & \multicolumn{1}{c|}{0.935} & \multicolumn{1}{c|}{0.938} & \multicolumn{1}{c|}{0.935} & \multicolumn{1}{c|}{193} & 7.30 \\ \hline
\begin{tabular}[c]{@{}c@{}}ResNet-50 (pre-trained)\end{tabular} & \multicolumn{1}{c|}{0.854} & \multicolumn{1}{c|}{0.862} & \multicolumn{1}{c|}{0.855} & 128 & \multicolumn{1}{c|}{0.830} & \multicolumn{1}{c|}{0.834} & \multicolumn{1}{c|}{0.831} & \multicolumn{1}{c|}{159} & 49.2 \\ \hline
\begin{tabular}[c]{@{}c@{}}MobileNet-V2 (pre-trained) \end{tabular} & \multicolumn{1}{c|}{0.880} & \multicolumn{1}{c|}{0.883} & \multicolumn{1}{c|}{0.881} & 83 & \multicolumn{1}{c|}{0.873} & \multicolumn{1}{c|}{0.934} & \multicolumn{1}{c|}{0.857} & \multicolumn{1}{c|}{\textbf{98}} & 2.58 \\ \hline
\begin{tabular}[c]{@{}c@{}}ConvNext-V2 (pre-trained) \end{tabular} & \multicolumn{1}{c|}{0.842} & \multicolumn{1}{c|}{0.849} & \multicolumn{1}{c|}{0.842} & 204 & \multicolumn{1}{c|}{0.834} & \multicolumn{1}{c|}{0.834} & \multicolumn{1}{c|}{0.830} & \multicolumn{1}{c|}{238} & 27.8 \\ \hline
\begin{tabular}[c]{@{}c@{}}ResNet-152$+$CBAM [36]\\(pre-trained)\end{tabular} & \multicolumn{1}{c|}{0.917} & \multicolumn{1}{c|}{0.919} & \multicolumn{1}{c|}{0.918} & 308 & \multicolumn{1}{c|}{0.894} & \multicolumn{1}{c|}{0.896} & \multicolumn{1}{c|}{0.895} & \multicolumn{1}{c|}{300} & 64.7 \\ \hline
\begin{tabular}[c]{@{}c@{}}Pooling-based ViT (PiT)\\ (pre-trained) [29]\end{tabular} & \multicolumn{1}{c|}{0.929} & \multicolumn{1}{c|}{0.930} & \multicolumn{1}{c|}{0.929} & 84 & \multicolumn{1}{c|}{0.915} & \multicolumn{1}{c|}{0.917} & \multicolumn{1}{c|}{0.915} & \multicolumn{1}{c|}{109} & 4.59 \\ \hline
\textbf{\begin{tabular}[c]{@{}c@{}}Proposed framework\\ (trained from scratch)\end{tabular}} & \multicolumn{1}{c|}{\textbf{0.943}} & \multicolumn{1}{c|}{\textbf{0.943}} & \multicolumn{1}{c|}{\textbf{0.943}} & \textbf{80} & \multicolumn{1}{c|}{\textbf{0.950}} & \multicolumn{1}{c|}{\textbf{0.951}} & \multicolumn{1}{c|}{\textbf{0.950}} & \multicolumn{1}{c|}{116} & \textbf{2.40} \\ \hline
\end{tabular}
}
\end{center}
\end{table*}

\begin{table}[t]
		\begin{center}		\caption{Testing results of proposed VGG-Lite$+$CEEM, for 5-fold cross validation on `Pneumonia Imbalance' dataset (\textbf{Weighted Average is displayed}). The best results and Mean $\pm$ Standard deviation value are represented in bold letters.}
            \vspace{0.2cm}
		\resizebox{1.01\columnwidth}{!}{
		\begin{tabular}{|c|c|c|c|c|c|}
			\hline
			{folds} & {Accuracy} & {Precision} & {Recall} & {F1-score} & {AUC}\\
                \hline
                \hline
			fold1  & 0.957 & 0.958 & 0.957 & 0.957 & \textbf{0.995}\\
			\hline
			fold2 & 0.948 & 0.949 & 0.947 & 0.948 & 0.991\\
			\hline
   fold3  & \textbf{0.958} & \textbf{0.959} & \textbf{0.957}& \textbf{0.958} & 0.994\\
			\hline
			fold4 & {0.931} & {0.932} & {0.929} & {0.931} & 0.989\\
   \hline
   fold5  & 0.941 & 0.946 & 0.938 & 0.942 & 0.991\\
			\hline
   \textbf{Mean $\pm$ Std} & \textbf{0.947}             & \textbf{0.949}             & \textbf{0.945}             & \textbf{0.947} & \textbf{0.992}\\
  \textbf{Deviation} & \textbf{$\pm$ 0.012} & \textbf{$\pm$ 0.013} & \textbf{$\pm$ 0.012} & \textbf{$\pm$ 0.012} & \textbf{$\pm$ 0.006} \\
   \hline
		\end{tabular}
  }
	\end{center}
\end{table}
This can be observed from TABLE-II that none of the trained-from-scratch existing models performed efficiently on both CXR datasets. The majority of these models failed to mitigate the class imbalance issue in those datasets, resulting in an accuracy not more than 88\%. In particular, Covid-Net and ViT models have suffered from slight overfitting due to deploying heavy model (means, having higher number of trainable parameters) in their models. Whereas, ``PneuNet" and ``Attention-based VGG-16" could not able to resolve class imbalance issue from these datasets. Specifically, ``Attention-based VGG-16" has severely failed on ``Pneumonia Imbalance dataset", resulting in only 47.4\% accuracy, and 27.9\% F1 score. Hence, these results presented in TABLE-II validate the fact that existing (trained-from-scratch models) were ineffective in directly addressing the class imbalance issue in these CXR datasets. This reveals that there is a significant research gap, as previously discussed in Section-II. From TABLE-II it is evident that the proposed ``VGG-Lite" and ``VGG-Lite$+$CEEM" frameworks outperformed all other existing models by substantial margins. Notably, the proposed framework ``VGG-Lite$+$CEEM" surpassed two recent state-of-the-art models, ViT and PneuNet (ResNet-18 + ViT), by approximately 10\% and 16\%, respectively, on the ``Pneumonia Imbalance Dataset", and by 10\% and 7\%, respectively, on the ``Original CXR Dataset".
Fig.5 illustrates a class-wise RoC graphs, obtained by those trained-from-scratch models. This can be noticed that for our proposed framework, the AUC values are consistently very close to 1 across all classes, however, for other existing models the AUC deviates slightly from 1 ($0.94$ to $0.99$).

Moreover, from TABLE-III, it is evident that all the pre-trained CNN models VGG-16 [42], MobileNet-V2 [43], Inception-V3 [49], DenseNet-121 [50], ResNet-50 [51], ConvNext-V2 [52] (i.e., 100\% fine-tuned from ImageNet dataset), have not resolved the class imbalance problem entirely from the challenging CXR datasets. Some of the models like ResNet-50, and ConvNext-V2 are prone to overfitting on both the datasets, due to the complexity (large number of layers) present in their model. Overall, only Inception-V3 has achieved 93\% accuracy consistently on both the datasets, among all pre-trained CNN models. This is to clarify that for a trained-from-scratch model, it is not very easy task to surpass the efficacy of all the pre-trained CNN models consistently. Nevertheless, our proposed model successfully achieved this milestone. It outperformed the latest pre-trained models Pooling-based ViT (PiT), ConvNext-V2 and ResNet-152$+$CBAM by substantial margins of 3.5\%, 11.6\%, and 5.5\%, respectively, in ``Pneumonia Imbalance Dataset". These are significant boosting performance. This boosting performance is prominent in other ``Original CXR Dataset" as well, which can be visualized from TABLE-III. Hence, TABLE-II and TABLE-III reveal that the proposed framework not only outperformed all the trained-from-scratch deep learning models, but also, demonstrated superior performance compared to all other pre-trained models. In particular, our proposed framework ``VGG-Lite$+$CEEM"  has achieved (weighted average of) 95\% accuracy, 95.1\% precision, 95\% recall, 95\% F1 score, and 99.4\% AUC on ``Pneumonia Imbalance Dataset''. This is the best result so far on this dataset. Nevertheless, it is observed from TABLE-II and TABLE-III that the number of trainable parameters, and time complexity (secs/ epochs) of the proposed ``VGG-Lite'' and ``VGG-Lite+CEEM'' models is considerably lesser than that of other existing CNN models. Only MobileNet-V2 and the pre-trained PiT model had similarly less parameters or model complexity compared to our proposed method. This proves that the proposed framework (or algorithm) not only offers superior efficacy compared to existing models, but also, exhibits minimal computational complexity, thereby making it adaptable for use on lightweight devices.

\subsection{Validity checking by 5-fold cross validation experiment}
For the validity purpose, we have also conducted a stratified 5-fold cross validation experiment on the ``Pneumonia Imbalance dataset". We effectively created the equivalent of 5 different datasets (we call them fold1-to-fold5 in TABLE-IV), where each dataset has distinct testing set, having different statistics compared to the same of other 4 datasets. In other words, we have created 5 different CXR datasets and we are seeking to compute the standard deviation of performances of the proposed framework on these 5 distinct datasets. Therefore, this environment is more challenging than the ``Pneumonia Imbalance dataset". We prefer stratified 5-fold cross-validation over stratified 10-fold cross-validation because, for minority classes like VP and TB, each fold may have an insufficient number of test samples, which may cause greater variability between the training and testing sets. The results of this 5-fold cross-validation, with mean and standard deviation values, are presented in TABLE-IV. From TABLE-IV it is evident that the proposed framework has attained a mean of 94.7\% accuracy, precision, recall and F1-score with standard deviation ($<=1.3\%$). This reveals that the proposed framework consistently performed on all the 5 folds of CXR datasets, with very less deviation. Hence, this experiment proves the validity or reliability of the proposed theory in this research. 

\section{Conclusion and Future Works}
A novel attention block ``Complementary and Edge Enhanced Module (CEEM)" was proposed in order to mitigate the class imbalance problem from CXR datasets. A ``Pneumonia Imbalance Dataset" was created and published on Kaggle, which was a more challenging and imbalanced dataset than the existing one. The proposed framework consisted of two main components, (I) VGG-Lite model was proposed as a base model which was a very lightweight model due to heavy utilization of DWSC layers. (II) CEEM attention block was introduced for the first time in which 2Max-Min pooling was incorporated. Unlike other pooling techniques, the 2Max-Min pooling layer had the ability to directly provide attention to the most salient features, that is, edges in the CXR datasets. As a consequence, the proposed framework (trained-from-scratch) alleviated the challenges from ``Pneumonia Imbalanced Dataset", as well as, it generalized well on other ``Original CXR dataset." The proposed framework ``VGG-Lite$+$CEEM" outperformed recent existing (trained-from-scratch) models and latest pre-trained CNN models, by significant margins on both datasets. Furthermore, a stratified 5-fold cross-validation experiment demonstrated a very low standard deviation ($<=1.3\%$) value of accuracy, and F1 score, across $5$ distinct folds. This exhibited a strong stability of the proposed model, thus, further validating our proposed theory.

\vspace{0.1cm} 
As part of our future work, we aim to extend this research by developing a universal standard deep learning architecture, referred to as \textit{Pneumonia-Net}, which will be capable of accurately detecting any variants of pneumonia across diverse Chest X-Ray and CT datasets. In this new standard \textit{Pneumonia-Net} model, Vision Transformer (ViT) encoder or a self supervised module could be leveraged in order to enhance the generalization ability of the proposed model, depending on the challenges of the datasets. Furthermore, we are also planning to collect more challenging and imbalanced clinical data (both CXR and CT) from hospitals to accomplish this future task. In this research, we did not yet deal with data affected by noise or artifacts, however, such issues are likely to arise in real-world clinical data. Therefore, in the future work, we plan to integrate a `problem-specific module' (just like CEEM) on top of the base model, in order to effectively handle noise-related challenges.


\begin{thebibliography}{1}

\bibitem{}
Hoare, Zara, and Wei Shen Lim. ``Pneumonia: update on diagnosis and management.'' \textit{Bmj}, vol. 332, no 7549, pp: 1077-1079, 2006.

\bibitem{}
Cozzi, Diletta, et al. ``Ground-glass opacity (GGO): a review of the differential diagnosis in the era of COVID-19." \textit{Japanese journal of radiology}, vol. 39, no. 8, pp: 721-732, 2021.
\bibitem{}
Sakula, Alex. ``Robert Koch: centenary of the discovery of the tubercle bacillus, 1882." \textit{Thorax}, vol. 37, no.4 pp: 246, 1982.
\bibitem{}
Dijkman, Karin, et al. ``Prevention of tuberculosis infection and disease by local BCG in repeatedly exposed rhesus macaques." \textit{Nature Medicine}, vol. 25, no. 2, pp: 255-262, 2019.
\bibitem{}
``The species Severe acute respiratory syndrome-related coronavirus: classifying 2019-nCoV and naming it SARS-CoV-2.'' \textit{Nature Microbiology}, vol. 5, no. 4, pp: 536-544, 2020.
\bibitem{}
Bartleson, Juliet M., et al. ``SARS-CoV-2, COVID-19 and the aging immune system.'' \textit{Nature Aging}, vol. 1, no.9, pp: 769-782, 2021.
\bibitem{}
Rossen, Lauren M., et al. ``Excess all-cause mortality in the USA and Europe during the COVID-19 pandemic, 2020 and 2021.'' \textit{Scientific Reports}, vol. 12, no.1, pp: 18559, 2022.
\bibitem{}
Bassi, Jessica, et al. ``Poor neutralization and rapid decay of antibodies to SARS-CoV-2 variants in vaccinated dialysis patients.'' \textit{PLoS One}, vol. 17, no.2, pp: e0263328, 2022.
\bibitem{}
Wang, Guangyu, et al. ``A deep-learning pipeline for the diagnosis and discrimination of viral, non-viral and COVID-19 pneumonia from chest X-ray images.'' \textit{Nature biomedical engineering}, vol. 5, no. 6, pp: 509-521, 2021.
\bibitem{}
Wang, Linda, Zhong Qiu Lin, and Alexander Wong. ``Covid-net: A
tailored deep convolutional neural network design for detection of covid-19 cases from chest x-ray images.” Scientific Reports, vol. 10, no.1, pp:
19549, 2020.
\bibitem{}
Chowdhury, Muhammad EH, et al. ``Can AI help in screening viral and
COVID-19 pneumonia?.” IEEE Access, vol. 8, pp: 132665-132676, 2020.
\bibitem{}
Kermany, Daniel S., et al. ``Identifying medical diagnoses and treatable diseases by image-based deep learning." \textit{cell}, vol. 172, no.5, pp: 1122-1131, 2018.
\bibitem{}
Rahman, Tawsifur, et al. ``Reliable tuberculosis detection using chest X-ray with deep learning, segmentation and visualization." \textit{IEEE Access}, vol. 8, pp: 191586-191601, 2020).
\bibitem{}
Tyagi, Mrinal et al., ``Custom Weighted Balanced Loss function for Covid 19 Detection from an Imbalanced CXR Dataset." \textit{26th International Conference on Pattern Recognition (ICPR) 2022}, pp: 2707-2713.
\bibitem{}
Roy, S. and Khurana, R. ``Mobile Freeze-Net with Attention-based Loss Function for Covid-19 Detection from an Imbalanced CXR Dataset." \textit{In Proceedings of the 38th ACM/SIGAPP Symposium on Applied Computing}, 2023, pp. 611-613.
\bibitem{}
Das, Dipayan, K. C. Santosh, and Umapada Pal. ``Truncated inception net: COVID-19 outbreak screening using chest X-rays." \textit{Physical and engineering sciences in medicine}, vol. 43, pp: 915-925, 2020.
\bibitem{}
Kumar, Aayush, et al. ``SARS-Net: COVID-19 detection from chest x-rays by combining graph convolutional network and convolutional neural network." \textit{Pattern Recognition}, vol. 122, pp: 108255, 2022.
\bibitem{}
Mamalakis, Michail, et al. ``DenResCov-19: A deep transfer learning network for robust automatic classification of COVID-19, pneumonia, and tuberculosis from X-rays." \textit{Computerized Medical Imaging and Graphics}, vol. 94, pp: 102008, 2021.
\bibitem{}
Ahmed, Syed Thouheed, et al. ``Augmented Intelligence Based COVID-19 Diagnostics and Deep Feature Categorization Based on Federated Learning.'' \textit{IEEE Transactions on Emerging Topics in Computational Intelligence}, 2024.
\bibitem{}
Roy, Santanu, et al. ``SVD-CLAHE boosting and balanced loss function for COVID-19 detection from an imbalanced Chest X-Ray dataset." \textit{Computers in Biology and Medicine}, vol. 150, pp: 106092, 2022.
\bibitem{}
Chamseddine, E., Mansouri, N., Soui, M., Abed, M. ``Handling class imbalance in covid-19 chest x-ray images classification: Using smote and weighted loss." \textit{Applied Soft Computing}, vol.129, pp: 109588, 2022.
\bibitem{}
Sun, Junding, et al. ``SAGCN: Self-adaptive Graph Convolutional Network for pneumonia detection." \textit{Biomedical Signal Processing and Control}, vol. 106, pp: 107634, 2025. 
\bibitem{}
Ali, Mudasir, et al. ``Pneumonia Detection Using Chest Radiographs With Novel EfficientNetV2L Model.'' \textit{IEEE Access}, 2024.
\bibitem{}
Guarrasi, V., D'Amico, N.C., Sicilia, R., Cordelli, E. and Soda, P., ``A multi-expert system to detect COVID-19 cases in X-ray images." \textit{In 2021 IEEE 34th International Symposium on Computer-Based Medical Systems (CBMS)}, IEEE, 2021, pp. 395-400. 
\bibitem{}
Zhou, Tao, Caiyue Peng, Yujie Guo, Hongxia Wang, Yuxia Niu, and Huiling Lu. ``Identity-Mapping ResFormer: A Computer-Aided Diagnosis Model for Pneumonia X-Ray Images." \textit{IEEE Transactions on Instrumentation and Measurement}, 2025.
\bibitem{}
Ranftl, R., Bochkovskiy, A., and Koltun, V. ``Vision transformers for dense prediction.'' \textit{In Proceedings of the IEEE/CVF international conference on computer vision}, 2021, pp. 12179-12188.

\bibitem{}
Jing, L., Tian, Y. ``Self-supervised visual feature learning with deep neural networks: A survey''. \textit{IEEE transactions on pattern analysis and machine intelligence}, vol.43, no.11, pp: 4037–4058, 2020. 
\bibitem{}
Liu, Z., Mao, H., Wu, C. Y., Feichtenhofer, C., Darrell, T., and Xie, S. ``A convnet for the 2020s''. \textit{In Proceedings of the IEEE/CVF conference on computer vision and pattern recognition}, 2022, pp. 11976-11986.
\bibitem{}
Heo, B., Yun, S., Han, D., Chun, S., Choe, J., Oh, S.J. ``Rethinking spatial dimensions of vision transformers.'' \textit{In: Proceedings of the IEEE/CVF International Conference on Computer Vision}, 2021, pp. 11936–11945.
\bibitem{}
Xie, Z., Lin, Y., Yao, Z., Zhang, Z., Dai, Q., Cao, Y., Hu, H. ``Self-supervised learning with swin transformers.'' arXiv preprint arXiv:2105.04553, 2021. 
\bibitem{}
Singh, Sukhendra, Manoj Kumar, Abhay Kumar, Birendra Kumar Verma, Kumar Abhishek, and Shitharth Selvarajan. ``Efficient pneumonia detection using Vision Transformers on chest X-rays." \textit{Scientific reports}, vol. 14, no. 1, pp: 2487, 2024. 
\bibitem{}
Wang, Tianmu et al., ``PneuNet: deep learning for COVID-19 pneumonia diagnosis on chest X-ray image analysis using Vision Transformer." \textit{Medical and Biological Engineering Computing}, vol. 61, no. 6, pp: 1395-1408, 2023.
\bibitem{}
Lin, Zhijie, et al. ``AANet: Adaptive attention network for COVID-19 detection from chest X-ray images.'' \textit{IEEE Transactions on Neural Networks and Learning Systems}, vol. 32, no. 11, pp: 4781-4792, 2021.
\bibitem{}
Mezina, Anzhelika, and Radim Burget. ``Detection of post-COVID-19-related pulmonary diseases in X-ray images using Vision Transformer-based neural network." \textit{Biomedical Signal Processing and Control}, vol. 87, pp: 105380, 2024. 
\bibitem{}
Hu, Jie, Li Shen, and Gang Sun. ``Squeeze-and-excitation networks.'' \textit{Proceedings of the IEEE conference on computer vision and pattern recognition}, 2018, pp: 7132- 7141.
\bibitem{}
Woo, S., Park, J., Lee, J. Y., and Kweon, I. S. ``Cbam: Convolutional block attention module'', \textit{In Proceedings of the European conference on computer vision (ECCV)}, 2018, pp. 3-19.
\bibitem{}
Xiao, Bin, et al. ``PAM-DenseNet: A deep convolutional neural network for computer-aided COVID-19 diagnosis." \textit{IEEE Transactions on Cybernetics}, vol. 52, no.11, pp: 12163-12174, 2021.
\bibitem{}
Sitaula, Chiranjibi, and Mohammad Belayet Hossain. ``Attention-based VGG-16 model for COVID-19 chest X-ray image classification." \textit{Applied Intelligence}, vol. 51, pp: 2850-2863, 2021.
\bibitem{}
Wang, Shui-Hua, et al. ``AVNC: attention-based VGG-style network for COVID-19 diagnosis by CBAM.'' \textit{IEEE Sensors Journal}, vol. 22, no. 18, pp: 17431-17438, 2021.
\bibitem{}
Singh, Sukhendra, Manoj Kumar, Abhay Kumar, Birendra Kumar Verma, and S. Shitharth. ``Pneumonia detection with QCSA network on chest X-ray." \textit{Scientific Reports}, vol. 13, no. 1, pp: 9025, 2023. 
\bibitem{}
Roy, Santanu, Archit Gupta, Shubhi Tiwari, and Palak Sahu. ``AD-Lite Net: A Lightweight and Concatenated CNN Model for Alzheimer’s Detection from MRI Images." \textit{In International Conference on Pattern Recognition (ICPR)}, Springer, Cham, 2025, pp. 1-16.
\bibitem{}
Simonyan, Karen, and Andrew Zisserman. ``Very deep convolutional networks for large-scale image recognition." \textit{arXiv preprint}, arXiv: 1409.1556, 2014.
\bibitem{}
A.G. Howard, M. Zhu, B. Chen, D. Kalenichenko, W. Wang, T. Weyand, M.
Andreetto, H. Adam, ``Mobilenets: Efficient convolutional neural networks for mobile vision applications," 2017, arXiv preprint arXiv:1704.04861.
\bibitem{}
Roy, Santanu, et al. ``A Novel Spatial Attention Module (SAM) for Alzheimer's Detection from MRI Images." \textit{In Proceedings of the Fifteenth Indian Conference on Computer Vision Graphics and Image Processing}, ACM, 2024, pp. 1-10.
\bibitem{}
Chollet, François. ``Xception: Deep learning with depthwise separable convolutions." \textit{Proceedings of the IEEE conference on computer vision and pattern recognition}, 2017.
\bibitem{}
Rahman, Tawsifur, et al. ``Exploring the effect of image enhancement techniques on COVID-19 detection using chest X-ray images." \textit{Computers in biology and medicine}, vol. 132, pp: 104319, 2021.
\bibitem{}
Wang, Zheng, et al. ``Automatically discriminating and localizing COVID-19 from community-acquired pneumonia on chest X-rays." \textit{Pattern recognition}, vol. 110, pp: 107613, 2021.
\bibitem{}
Roy, Santanu, Ashvath Suresh, and Archit Gupta. ``Edge Attention Module for Object Classification." 2025, arXiv preprint arXiv:2502.03103.
\bibitem{}
C. Szegedy, V. Vanhoucke, S. Ioffe, J. Shlens, Z. Wojna, ``Rethinking the inception architecture for computer vision'', \textit{In: Proceedings of the IEEE Conference on Computer Vision and Pattern Recognition}, 2016, pp. 2818–2826.
\bibitem{}
G. Huang, Z. Liu, L. Van Der Maaten, K.Q. Weinberger, ``Densely connected
convolutional networks,'' \textit{In: Proceedings of the IEEE Conference on Computer Vision and Pattern Recognition}, 2017, pp. 4700–4708. 
\bibitem{}
K. He, X. Zhang, S. Ren, J. Sun, ``Deep residual learning for image recognition,'' \textit{In: Proceedings of the IEEE Conference on Computer Vision and Pattern Recognition}, 2016, pp. 770–778. 
\bibitem{}
Woo et al., ``Convnext v2: Co-designing and scaling convnets with masked autoencoders." \textit{In Proceedings of the IEEE/CVF conference on computer vision and pattern recognition}, 2023, pp. 16133-16142.

\end{thebibliography}
\end{document}